\documentclass{ws-ijmpcs}
\usepackage{float}

\begin{document}

\markboth{Masheng Yang \& Yaping Cheng}
{Energy Non-linearity studies at Daya Bay}

\catchline{}{}{}{}{}
%
%%%%%%%%%%%%%%%%%%%%%%%%%%%%%%%%%%%%%%%%%%%%%%%%%%%%%%%%%%%%%%%%%%%%

\title{Energy Non-linearity Studies at Daya Bay}

\author{Masheng Yang}
\address{Institute of High Energy Physics, Beijing, China  \\
%\address{University of Chinese Academy of Sciences, Beijing, China\\
yangms@ihep.ac.cn}
\author{Yaping Cheng}
\address{Institute of High Energy Physics, Beijing, China \\
%\address{University of Chinese Academy of Sciences, Beijing, China\\
chengyp@ihep.ac.cn}

\maketitle

\begin{center}
{\itshape On behalf of the Daya Bay Collaboration}
\end{center}

%\begin{history}
%\received{Day Month Year}
%\revised{Day Month Year}
%\end{history}

\begin{abstract}

     The Daya Bay Reactor Neutrino Experiment has measured a non-zero value of the neutrino
mixing angle $\theta_{13}$ with a significance of 7.7 standard deviations by a rate-only analysis\cite{dybcpc}.The distortion of neutrino energy spectrum carries additional oscillation information and can improve the sensitivity of $\theta_{13}$ as well as measure neutrino mass splitting $\Delta m^{2}_{ee}$. A rate plus shape analysis is performed and the results have been published\cite{spectraana}.Understanding detector energy non-linearity response is crucial for the rate plus shape analysis. In this contribution, we present a brief description of energy non-linearity studies at Daya Bay.

\end{abstract}

\ccode{PACS numbers:}

\section{General introduction}
    It is well established that the flavor of a neutrino oscillates
with time. Neutrino oscillations can be described by the
three mixing angles($\theta_{12}$, $\theta_{23}$, and $\theta_{13}$,),
 a phase of the Pontecorvo-Maki-Nakagawa-Sakata matrix, and
the neutrino mass squared differences\cite{cite1}$^{,}$\cite{cite2}.
To date, the Daya Bay experiments has made the most precise measurement of the neutrino mixing angle
$\theta_{13}$\cite{dybcpc}$^{,}$\cite{dybprl}. The Daya Bay Experiment
 has three underground Experiment Halls(EH) and totally 8 anti-neutrino detectors(ADs).
 The AD contains a structure of three layers, with Gadolinium loaded Liquid Scintillator(GdLS) in the center, LS in the
 middle as the
gamma catcher, and oil in the outer layer to shield the radioactive components like PMTs.
 The $\overline{\nu}_{e}$ from the reactor interacts with the detector via the inverse beta decay(IBD),
and a positron together with a neutron come out after the interaction, namely, $\overline{\nu}_{e}+p\rightarrow e^{+}+n$.
With the positron
kinetic energy  deposited followed by its annihilation, this event presents as a prompt signal. The neutron
is captured by the Gd nucleus, and several gammas with total energy about 8 MeV are emmited.
 This event forms the delayed signal. IBD candidates are selected
through time-correlation method. The prompt energy from the positron
gives an estimate of the incident $\overline{\nu}_{e}$ energy through: $E_{\nu}\simeq E_{e^{+}}+0.8MeV$.

\section{Energy response process}

    The Daya Bay detector energy response can be understood as follows. A particle with its true energy deposit its
energy in the AD. For a $e^{-}$, $E_{true}$ is the kinetic energy; for a positron, $E_{true}$ is
the sum of the kinetic energy and the energy from annihilation. After that, the LS translates the deposit energy into
visible energy $E_{vis}$. The visible photons are detected by photomultiplier tubes(PMT).
 After the calibration and reconstruction, $E_{vis}$ is converted to be the reconstructed energy $E_{rec}$.
The energy response is not linear due to scintillator and electronics effects.
 The non-linearity, $E_{rec}$/$E_{true}$,
can be separated into two parts: the electronics non-linearity, $E_{rec}$/$E_{vis}$, and the scintillator non-linearity,
$E_{vis}$/$E_{true}$. Non-linearity study is significant to deduce the true energy of positron from the reconstructed
energy.

\section{Non-linearity from the electronics}

\begin{figure}[H]
\centerline{\includegraphics[width=4.7cm]{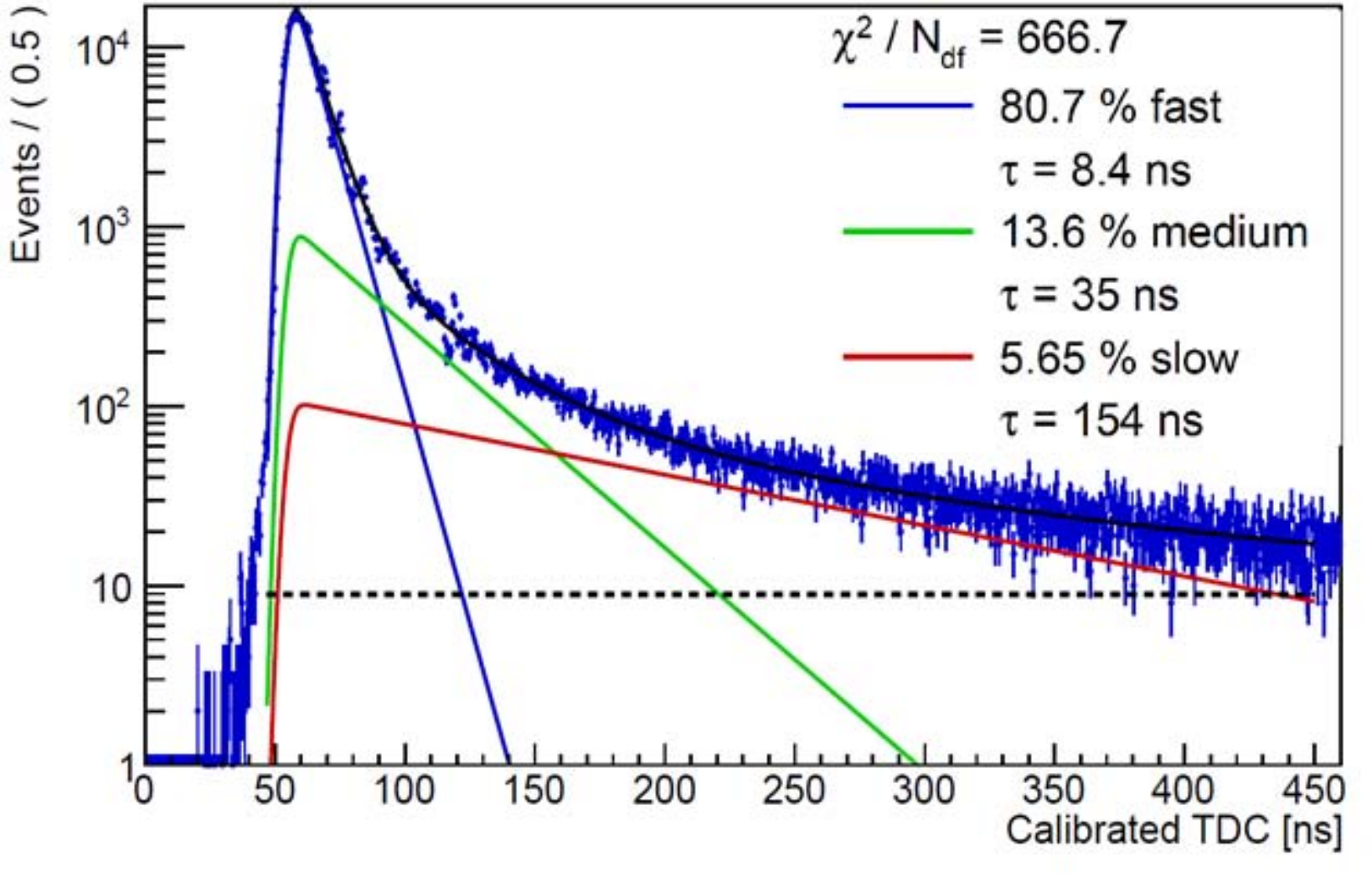}}
\vspace*{8pt}
\caption{Fast and slow components of LS photons, fitted by exponential functions. \label{elecnl}}
\end{figure}

    The electronics non-linearity is due to the interplay between slow component of LS light emitting $>$ 100 ns later after the first light(see Fig.~\ref{elecnl})
      and the front end electronics system.
      Later hits formed by slow component may not be included in the hit collection, thus would make the
      collection efficiency decrease with increasing $E_{vis}$. We used several models to
      parameterize the electronics non-linearity, for example,
      $\frac{E_{vis}}{E_{rec}}=(1-\alpha e^{\frac{E_{rec}}{\tau}})$.

\section{Non-linearity from liquid scintillator}

\begin{figure}[H]
\centerline{\includegraphics[width=4.7cm]{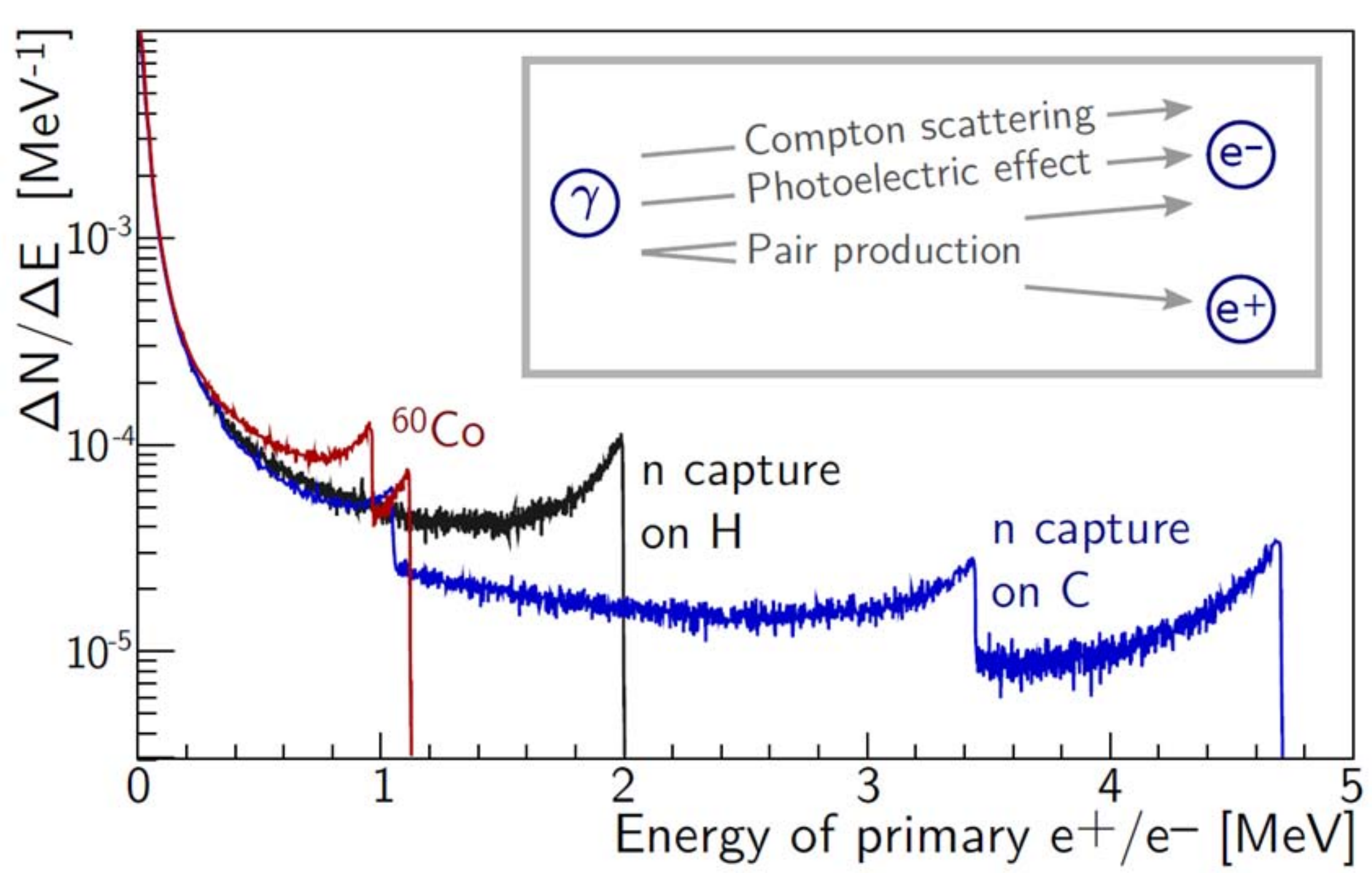}}
\vspace*{8pt}
\caption{The gamma to electron conversion probability density function via Geant4 for different gammas. \label{gamma2e}}
\end{figure}

The scintillator nonlinearity is particle and energy-dependent, and it is related to the intrinsic scintillator quenching
and Cherenkov light emission.
Different models are used to constrain the LS non-linearity of electron. One is a model using the
Birk's law\cite{birkslaw} and Cherenkov radiation theory, $E_{vis}/E_{true}=f_{q}(E_{true};K_{B})+K_{C}，f_{c}(E_{true})$,
with the 1st term for quenching effect using the Birk's law, and the 2ns term for Cherenkov radiation. Another one is an
empirical model with 4 parameters, $E_{vis}/E_{true}=(p_{0}+p_{3}，E_{true})/(1+p_{1}，e^{-p_{2}，E_{true}})$. The gamma
non-linearity and
 the electron non-linearity in the LS are connected by the energy conversion processes,
  gamma-rays interact with matter mainly in three ways, namely, Computon scattering,Photoelectric, and
pair production, all these finally result in $e^{+}$ or $e^{-}$.
With a Geant4 simulation, the gamma to $e^{+}/e^{-}$ converting probability function
can be obtained.
 The non-linearity of gamma can be deduced
 from the secondary electron's non-linearity(see Fig.~\ref{gamma2e}).

\section{Available data to constrain the non-linearity}
\begin{figure}[H]
%\centerline{\includegraphics[width=4.7cm,height=3cm]{gammapoints}\includegraphics[width=4.7cm,height=3cm]{B12}}

\centerline{\includegraphics[height=3cm]{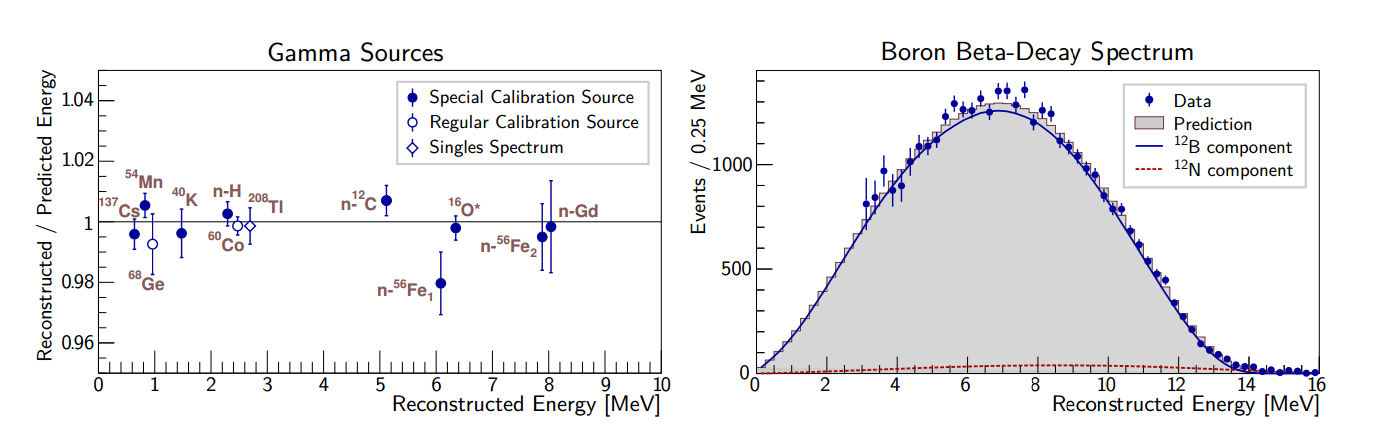}}
\vspace*{8pt}
\caption{The comparison of predicted vs measure gamma peaks(left) and $^{12}B$ spectrum (right). \label{B12}}
\end{figure}

During a special calibration period in summer 2012, we used gamma sources and neutron sources
 deployed at the center of detector
to do the nonlinearity study. Gamma sources such as $^{137}Cs,^{54}Mn,^{40}K$, and neutron sources such as
$^{241}Am-^{9}Be$ and $Pu-^{13}C$ were used. Neutrons emitted from the neutron radiation sources can be captured on
hydrogen or Gd. A single ~2.2MeV gamma is emitted when captured on hydrogen,
 providing one more gamma peak for nonlinearity study.
The calibration data were fitted with the models described above, and consistency is obtained among different models.
Energy spectrum from
  cosmic muon induced $\gamma+^{12}B$ spectrum, is also used to test the consistency of the nonlinearity. The comparisons
  of the predicted using the best fit nonlinearity model
vs measured gamma peaks and $^{12}B$ spectrum are shown in Fig.\ref{B12}.

\section{Summary}
\begin{figure}[H]
\centerline{\includegraphics[width=4.7cm,height=3cm]{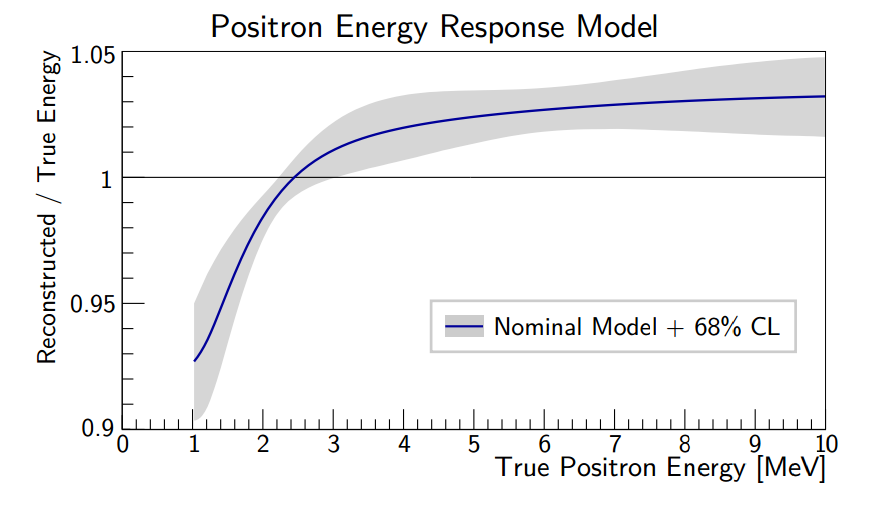}}%
\vspace*{8pt}
\caption{Positron non-linearity result and its uncertainty contour. \label{result}}
\end{figure}

	After comparison between different models and comparison between models and data,
 the energy non-linearity for positron is show as
follow(see Fig.~\ref{result}), the shadow area represents the nonlinearity uncertainty both
systematic and statistic combined within 1 sigma significance level,
 and the relative uncertainty of the non-linearity for
positron is about 1.5\%.

\appendix

\section*{Acknowledgments}
%
%This section should come before the References. Dedications and funding
%information may also be included here.
We'd like to thank the PIC organization committee for this opportunity to present the nonlinearity studies at Daya Bay. We
would also like to thank everyone who helped us on this contribution.

%\begin{thebibliography}{000} %for 3 digits
%\begin{thebibliography}{00}  %for 2 digits

%\begin{figure}[pb]
%\centerline{\includegraphics[width=4.7cm]{ijmpcsf1}}
%\vspace*{8pt}
%\caption{A schematic illustration of dissociative recombination. The
%direct mechanism, 4m$^2_\pi$ is initiated when the
%molecular ion S$_{\rm L}$ captures an electron with
%kinetic energy. \label{f1}}
%\end{figure}

\end{document}